\documentclass[12pt,english,english]{article}
\usepackage[T1]{fontenc}
\usepackage[latin1]{inputenc}
\usepackage{babel}
\usepackage{graphics}

\makeatletter

\providecommand{\LyX}{L\kern-.1667em\lower.25em\hbox{Y}\kern-.125emX\@}

 \newcommand{\lyxaddress}[1]{
   \par {\raggedright #1 
   \vspace{1.4em}
   \noindent\par}
 }

\usepackage[T1]{fontenc}
\usepackage[latin1]{inputenc}
\usepackage{amsmath}
\usepackage{amssymb}

\makeatletter

\usepackage[T1]{fontenc}
\usepackage[latin1]{inputenc}
\makeatletter

\usepackage{graphicx}

\def\fnum@table{\tablename~{\bf\thetable}}
\def\fnum@figure{\figurename~{\bf\thefigure}}
\def\tablename{\footnotesize{\bf Table}}
\def\figurename{\footnotesize{\bf Figure}}

\usepackage{babel}
\makeatother

\usepackage{babel}
\makeatother

\makeatother
\begin{document}

\title{Micro-Canonical Hadron Production in pp collisions }

\author{F. M. Liu\( ^{1,2} \)%
\thanks{Fellow of Alexander von Humboldt Foundation 
}, J. Aichelin\( ^{3} \), M. Bleicher\( ^{2} \) , K. Werner\( ^{3} \)}

\maketitle

\lyxaddress{\( ^{1} \)Institute of Particle Physics, Central China Normal University,
Wuhan, China\\
 \( ^{2} \)Institut f\"{u}r Theoretische Physik, J.W.Goethe Universit\"{a}t,
Frankfurt am Main, Germany\\
 \( ^{3} \)Laboratoire SUBATECH, University of Nantes - IN2P3/CNRS
- Ecole des Mines de Nantes, Nantes, France}

\begin{abstract}
We apply a microcanonical statistical model to investigate hadron
production in pp collisions. The parameters of the model are the energy
\( E \) and the volume \( V \) of the system, which we determine
via fitting the average multiplicity of charged pions, protons and
antiprotons in pp collisions at different collision energies. We then
make predictions of mean multiplicities and mean transverse momenta
of all identified hadrons. Our predictions on nonstrange hadrons are
in good agreement with the data, the mean transverse momenta of strange
hadron as well. However, the mean multiplicities of strange hadrons
are overpredicted. This agrees with canonical and grandcanonical studies,
where a strange suppression factor is needed. We also investigate
the influence of event-by-event fluctuations of the \( E \) parameter.

\vspace{.6cm}
\end{abstract}

\section{Introduction}

In pp collisions at high energies a multitude of hadrons is produced.
In contradistinction to the pp collisions at low energies even effective
theories are not able to provide the matrix elements for these reactions
and therefore a calculation of the cross section is beyond the present
possibilities of particle physics. In addition, even at moderate energies,
many different particles and resonances may be created and therefore
the number of different final states becomes huge.

In this situation, statistical approaches may be of great help \cite{Fermi:1950jd,Landau}.
It was Hagedorn who noticed that the transverse mass distributions
in high energy hadron-hadron collisions show a common slope for all
observed particles\cite{hag1}. This may be interpreted as a strong
hint that it is not the individual matrix elements but phase space
who governs the reaction. Therefore Hagedorn introduced statistical
methods into the strong interaction physics in order to calculate
the momentum spectra of the produced particles and the production
of strange particles.

Later, after statistical models have been successfully applied to
relativistic heavy ion collisions \cite{Hagedorn,Siemens,Mekjian,Csernai,Stoecker,Cleymans,Rafelski,Braun-Munzinger},
Becattini and Heinz \cite{bec} came back to the statistical description
of elementary \( pp \) and \( \bar{p}p \) reactions and used a canonical
model (in which the multiplicity of hadrons M is a function of volume
and temperature M(V,T)) in order to figure out whether the particle
multiplicities predicted by this approach are in agreement with the
(in the meantime very detailed) experimental results. For a center
of mass energy of around 20 GeV they found for non strange particles
a very good agreement between statistical model predictions and data
assuming that the particles are produced by a hadronic fireball with
a temperature of T = 170 MeV. The strange particles, however, escaped
from this systematics being suppressed by factors of the order of
two to five. Becattini and Heinz coped with this situation by introducing
a \( \gamma _{S} \) factor into the partition sum which was adjusted
to reproduce best the multiplicity of strange particles as well.

Statistical models are classified according to the implementation
of conservation laws:

\begin{itemize}
\item microcanonical: both, material conservation laws (\( Q, \) \( B \),
\( S, \) \( C, \) \( \cdots  \)) and motional conservation laws
(\( E \), \( \overrightarrow{p} \),\( \overrightarrow{J} \), \( \cdots  \)),
hold exactly.
\item canonical: material conservation laws hold exactly, but motional conservation
laws hold on average (a temperature is introduced).
\item grand-canonical: both material conservation laws and motional conservation
laws hold on the average (temperature and chemical potentials introduced).
\end{itemize}
The intensive physical quantities such as particle density and average
transverse momentum are independent of volume in the grand-canonical
calculation , while they depend on volume in both canonical and microcanonical
calculations. What one naively expects is that the microcanonical
ensemble must be used for very small volumes, for intermediate volumes
the canonical ensemble should be a good approximation, while for very
large volumes the grand-canonical ensemble can be employed. A numerical
study of volume effects in paper\cite{fu} tells us how big the volumes
need to be in order to make the grand-canonical ensembles applicable.
The comparison between the microcanonical and the canonical treatment
in paper\cite{fu} shows a very good agreement in particle yields,
when the same volume and energy density are used, and the strangeness
suppression is canceled in the canonical calculation.

In this paper, first we ignore the fluctuations of microcanonical
parameters and try to fix the microcanonical parameters, energy \( E \)
and volume \( V \), from fitting \( 4\pi  \) yields of protons,
antiprotons and charged pions from pp collisions. The one-to-one relation
between the collision energy \( \sqrt{s} \) and a pair of microcanonical
parameters \( E \) and \( V \) makes a link between the pp experiments
and the microcanonical approaches (or more generally, the statistical
ensembles). One can easily judge if grand-canonical ensembles can
describe pp collisions at any given energy; one can also transform
the fitting results to the canonical case and find the corresponding
temperature and volume of pp collisions at any energy.

Then we study the effect from the fluctuations of the microcanonical
energy parameter at a collision energy of \( 200\, \mathrm{GeV} \),
to check how reliable it is to fix microcanonical parameters without
energy fluctuations.

Finally, we would like to make a comparison between statistical models
and string models in describing pp collisions. This microcanonical
model and this fitting work will provide us a bridge to compare the
two classes of models and help us to understand the reaction dynamics.
In principle, one can consider a string as an ensemble of fireballs,
which may be considered as one effective fireball, when only total
multiplicities are considered.

\section{The approach}

We consider the final state of a proton-proton collision as a {}``cluster''
, {}``droplet'' or {}``fireball'' characterized by its volume
\( V \) (the sum of individual proper volumes), its energy \( E \)
(the sum of all the cluster masses) and the net flavour content \( Q=(N_{u}-N_{\bar{u}},N_{d}-N_{\bar{d}},N_{s}-N_{\bar{s}}) \),
decaying {}``statistically'' according to phase space. More precisely,
the probability of a cluster to hadronize into a configuration \( K=\{h_{1},p_{1};\ldots ;h_{n},p_{n}\} \)
of hadrons \( h_{i} \) with four momenta \( p_{i} \) is given by
the micro-canonical partition function \( \Omega (K) \) of an ideal,
relativistic gas of the \( n \) hadrons \cite{wer},\[
\Omega (K)=\frac{V^{n}}{(2\pi \hbar )^{3n}}\, \prod _{i=1}^{n}g_{i}\, \prod _{\alpha \in \mathcal{S}}\, \frac{1}{n_{\alpha }!}\, \prod _{i=1}^{n}d^{3}p_{i}\, \delta (E-\Sigma \varepsilon _{i})\, \delta (\Sigma \vec{p}_{i})\, \delta _{Q,\Sigma q_{i}},\]
 with \( \varepsilon _{i}=\sqrt{m_{i}^{2}+p_{i}^{2}} \) being the
energy, and \( \vec{p}_{i} \) the 3-momentum of particle \( i \).
The term \( \delta _{Q,\Sigma q_{i}} \) ensures flavour conservation;
\( q_{i} \) is the flavour vector of hadron \( i \). The symbol
\( \mathcal{S} \) represents the set of hadron species considered:
we take \( \mathcal{S} \) to contain the pseudoscalar and vector
mesons \( (\pi ,K,\eta ,\eta ',\rho ,K^{*},\omega ,\phi ) \) and
the lowest spin-\( \frac{1}{2} \) and spin-\( \frac{3}{2} \) baryons
\( (N,\Lambda ,\Sigma ,\Xi ,\Delta ,\Sigma ^{*},\Xi ^{*},\Omega ) \)
and the corresponding antibaryons. \( n_{\alpha } \) is the number
of hadrons of species \( \alpha  \), and \( g_{i} \) is the degeneracy
of particle \( i \). We generate randomly configurations \( K \)
according to the probability distribution \( \Omega (K) \). For the
details see ref. \cite{wer}. The Monte Carlo technique allows to
calculate mean values of observables as\[
\bar{A}=\sum _{K}A(K)\, \Omega (K)/\sum _{K'}\Omega (K'),\]
 where \( \sum  \) means summation over all possible configurations
and integration over the \( p_{i} \) variables. \( A(K) \) is some
observable assigned to each configuration, as for example the number
\( M_{h}(K) \) of hadrons of species \( h \) present in \( K \).
Since \( \bar{A} \) depends on \( E \) and \( V \), we usually
write \( \bar{A}(E,V) \). \( Q \) is not mentioned, since we only
study \( pp \) scattering here, therefore \( Q \) is always \( (4,2,0) \).

Let us consider the hadron multiplicity \( \bar{M}_{h}(E,V) \). This
quantity is used to determine the energy \( E(\sqrt{s}) \) and the
volume \( V(\sqrt{s}) \) which reproduces best the measured multiplicity
of some selected hadrons in pp collisions at a given \( \sqrt{s} \)
. This is achieved by minimizing \( \chi ^{2} \): \[
\chi ^{2}(E,V)=\frac{1}{\alpha }\sum _{j=1}^{\alpha }\frac{[\bar{M}_{\mathrm{exp},j}(\sqrt{s})-\bar{M}_{j}(E,V)]^{2}}{\sigma _{j}^{2}}\]
 where \( \bar{M}_{\mathrm{exp},j}(\sqrt{s}) \) and \( \sigma _{j} \)
are the experimentally measured multiplicity and its error of the
particle species \( j \) in pp collisions at an energy of \( \sqrt{s} \).

We start out our investigation by taking as input the most copiously
produced particles (\( j=p,\bar{p},\pi ^{+},\pi ^{-} \)). The data
have been taken from \cite{gia}. Whenever the data are not available,
the extrapolation of multiplicities by Antinucci \cite{gia} is used.
Fig.1 displays the results of our fit procedure in comparison with
the experimental data. We observe that these 4 particle species can
be quite well described by a common value of \( E(\sqrt{s}) \) and
\( V(\sqrt{s}) \).

\begin{figure}
{\centering \resizebox*{0.8\textwidth}{!}{\includegraphics{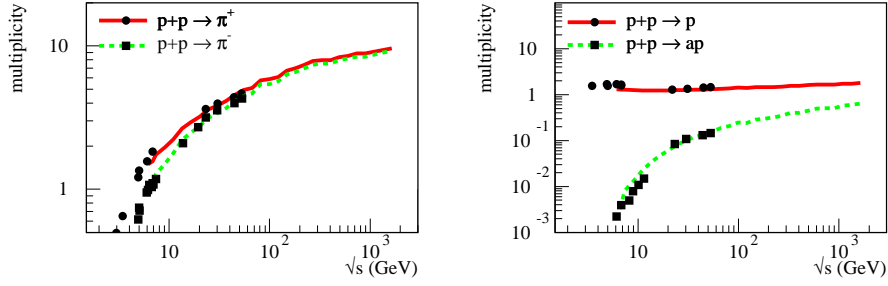}} \par}

\caption{The 4\protect\( \pi \protect \) multiplicities of \protect\( \pi ^{+},\pi ^{+}\protect \),
proton, antiproton produced in a pp collision as a function of \protect\( \sqrt{s}\protect \).
The full and dashed lines show the result of the \protect\( \chi ^{2}\protect \)
fit. The points are data from \cite{gia}. }
\end{figure}

Fig.2 shows \( E(\sqrt{s}) \) and \( V(\sqrt{s}) \), and fig.3 the
energy density \( \epsilon (\sqrt{s})=E(\sqrt{s})/V(\sqrt{s}) \),
which we obtain as the result of our fit. Both energy and volume increase
with \( \sqrt{s} \) but rather different, as the energy density shows.
We parameterize the energy and volume dependence on the collision
energy \( \sqrt{s} \) in eq. (\ref{eq:ev}). \begin{eqnarray}
E/\mathrm{G}eV & = & -3.8+3.76\mathrm{ln}\sqrt{s}+6.4/\sqrt{s}\nonumber \\
V/\mathrm{f}m^{3} & = & -30.0376+14.93\mathrm{ln}\sqrt{s}-0.013\sqrt{s}\label{eq:ev} 
\end{eqnarray}
 where \( \sqrt{s} \) is in unit GeV. Below \( \sqrt{s} \) = 8 GeV
the fit produces volumes below 2fm\( ^{3} \) which cannot be interpreted
physically. Above \( \sqrt{s} \) = 8 GeV the volume increases very
fast as compared to the energy giving rise to a decrease in the energy
density until - around \( \sqrt{s} \) = 200 GeV - the expected saturation
sets in and the energy density becomes constant. In view of the large
volume observed for these large energies the density of the different
particles does not change anymore \cite{fu} and therefore the particle
ratios stay constant above this energy.

\begin{figure}
{\centering \resizebox*{0.8\textwidth}{!}{\includegraphics{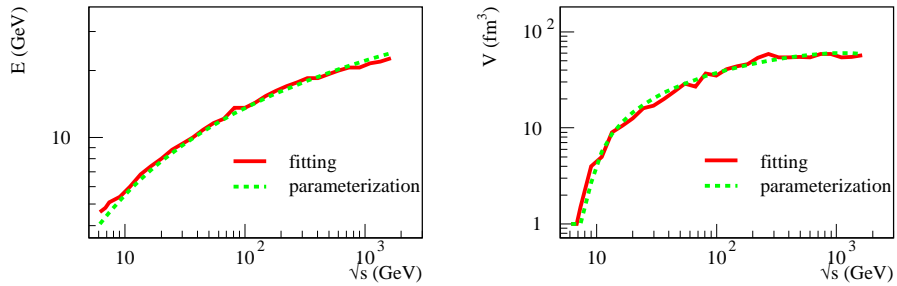}} \par}

\caption{The dependence of micro-canonical parameters \protect\( E\protect \)
(left) and volume \protect\( V\protect \) (right) on the collision
energy \protect\( \sqrt{s}\protect \) . The parameterization described
as eq. (\ref{eq:ev}) is plotted as dashed lines.}
\end{figure}

\begin{figure}
{\centering \resizebox*{0.4\textwidth}{!}{\includegraphics{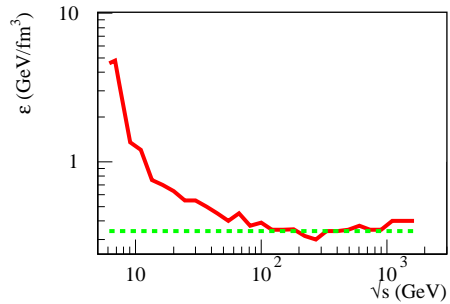}} \par}

\caption{The dependence of energy density \protect\( \varepsilon =E/V\protect \)
on the collision energy \protect\( \sqrt{s}\protect \). The dashed
line corresponds to constant energy density \protect\( 0.342\, \mathrm{G}eV/\mathrm{f}m^{3}\protect \)
which comes from a canonical calculation\cite{bec,fu}. }
\end{figure}

The quality of the fit can be judged from fig.4 where we have plotted
the \( \chi ^{2} \) values obtained for different values of \( E \)
and \( V \) and for \( \sqrt{s} \) = 200 GeV. We see that the energy
variation is quite small whereas the volume varies more. Nevertheless
the energy density is rather well defined.

\begin{figure}
{\centering \resizebox*{0.8\textwidth}{!}{\includegraphics{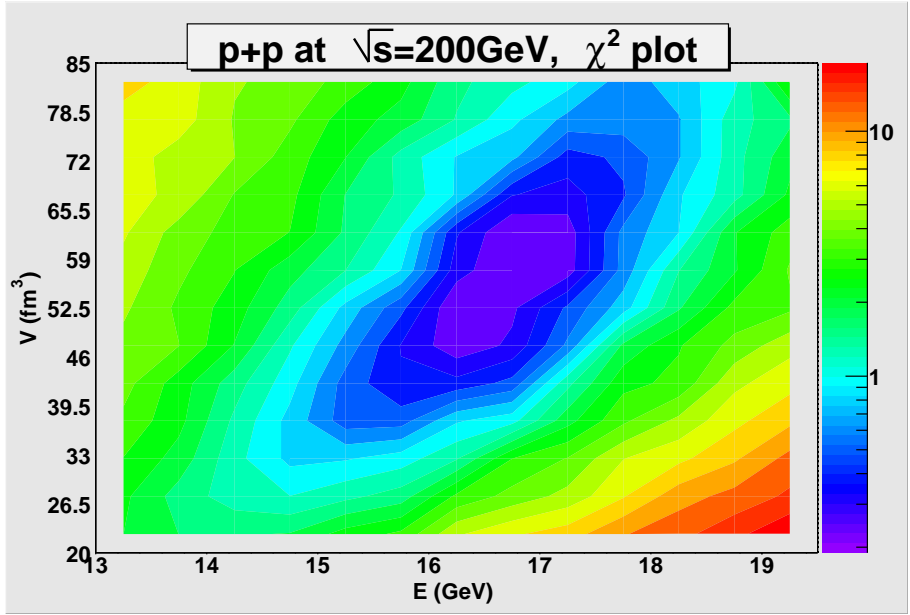}} \par}

\caption{The value of \protect\( \chi ^{2}\protect \) for different values
of \protect\( E\protect \) and \protect\( V\protect \) for a pp
reaction at \protect\( \sqrt{s}=200\, \mathrm{GeV}\protect \).}
\end{figure}

After having fitted the \( E(\sqrt{s}) \) and \( V(\sqrt{s}) \)
in using \( p,\bar{p},\pi ^{+} \) and \( \pi ^{-} \) data we can
now use these fitted values to predict the multiplicity of other hadrons.
This study we start in Fig. 5, where we present the multiplicity of
\( \pi ^{0} \) and \( \rho ^{0} \). For these particles experimental
data are available. We see that the absolute value as well as the
trend of the experimental data is quite well reproduced. The result
for those hadrons, for which no or only few data are available, is
displayed in fig. 6. As one can see the overall agreement is remarkable.
We would like to mention that we have as well made a \( \chi ^{2} \)
fit using as input the measured multiplicities of \( p,\bar{p} \)
and \( \rho ^{0} \). The results for \( E(\sqrt{s}) \) and \( V(\sqrt{s}) \)
differ only marginally.
\begin{figure}
{\centering \resizebox*{0.4\textwidth}{!}{\includegraphics{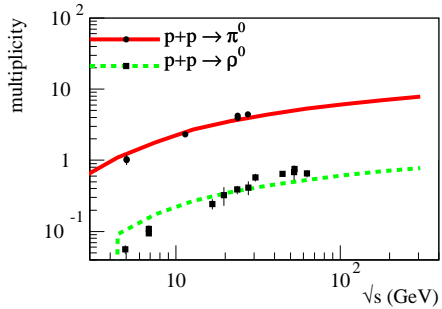}} \par}

\caption{Prediction of the \protect\( \pi ^{0}\protect \)and \protect\( \rho ^{0}\protect \)
multiplicity in pp collisions as a function of \protect\( \sqrt{s}\protect \).
The result of the calculation is compared to the data \cite{rho}.}
\end{figure}

\begin{figure}
{\centering \resizebox*{0.8\textwidth}{!}{\includegraphics{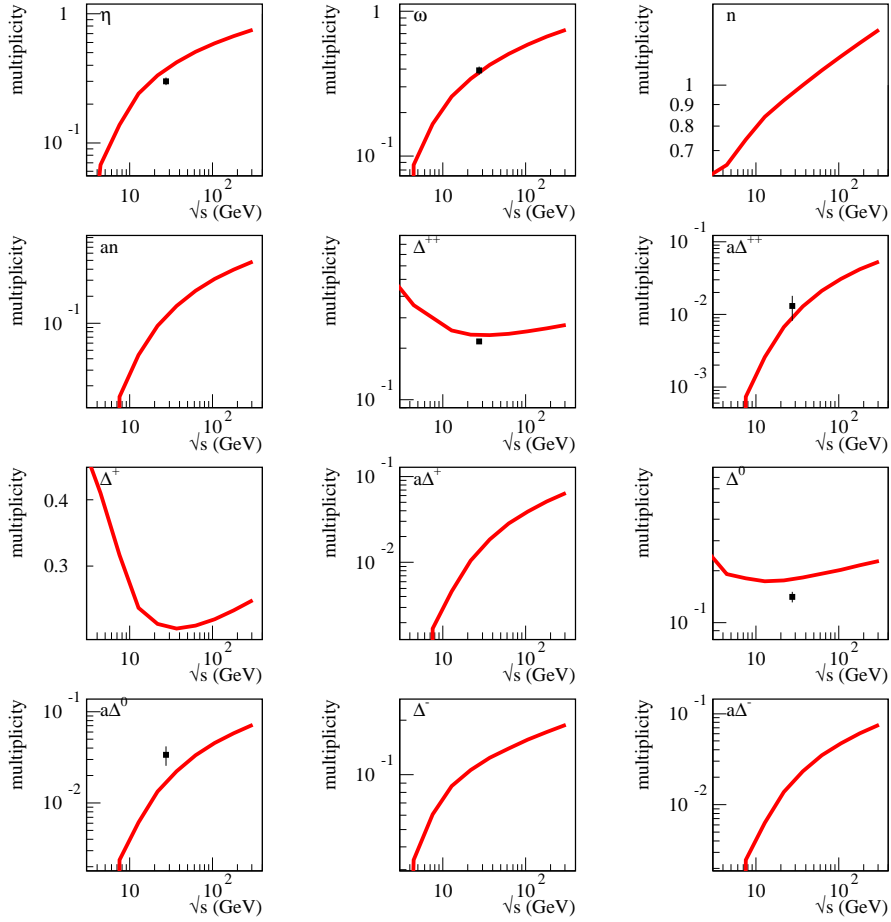}} \par}

\caption{Predictions of the multiplicities of non strange hadrons in pp collisions
as a function of \protect\( \sqrt{s}\protect \). We have plotted,
if available, also the data points for \protect\( \sqrt{s}=27.5\, \mathrm{GeV}\protect \)~\cite{pp}}
\end{figure}

\section{Strange particles}

With the parameters \( E(\sqrt{s}) \) and \( V(\sqrt{s}) \) which
we have obtained from the fit of the \( p,\, \bar{p},\, \pi ^{+} \)
and \( \pi ^{-} \) multiplicities, we can as well calculate the multiplicity
of strange particles or particles with hidden strangeness. The results
of these fits are presented in Fig. 7. As we can see immediately the
results for those particles are not at all in agreement with the data.
\( \Lambda  \) and \( \phi  \) multiplicities are off by a factor
of 3-5 roughly, for the \( \bar{\Lambda } \) the situation may be
similar but the spread of the experimental data does not allow for
a conclusion yet. Only the kaons come closer to the experimental values.
Although at lower energies a part of this deviation may come from
the fact that in our Monte Carlo procedure weak decays are neglected
and therefore \( K^{0} \) and \( \bar{K}^{0} \) are the particle
states which are treated, at higher energies this is not of concern
anymore. At \( \sqrt{s}=53\, \mathrm{GeV} \), we find - as in experiment
- that \( K_{L}=K_{S} \) and hence \( K^{0}=\bar{K}^{0} \). Therefore,
as in experiment, one finds that the strangeness contained in \( \Lambda ,\, \bar{\Lambda },\, K^{+} \)
and \( K^{-} \) adds up to zero. The absolute numbers are, however,
rather different: experimentally one finds .41 \( K^{+} \), 0.29
\( K^{-} \) and 0.12 \( \Lambda  \) \cite{ros,tab}, whereas the
fit yields 1.16 \( K^{+} \), 0.66 \( K^{-} \) and 0.58 \( \Lambda  \).

One is tempted to try to fit the strange particles separately. The
result at large \( \sqrt{s} \) is that, in contradiction to experiment,
more \( K^{0} \) then \( \bar{K}^{0} \) are produced. Consequently,
the strangeness in \( \Lambda ,\, \bar{\Lambda },\, K^{+} \) and
\( K^{-} \) does not add up to zero and the fit is far away from
the data. Thus we have to conclude that the strange particle multiplicities
cannot be described in a phase space approach using the parameters
one obtains from the fit of non strange particles, and that there
is no understanding presently why the suppression factor is rather
different for the different hadrons.
\begin{figure}
{\centering \resizebox*{0.8\textwidth}{!}{\includegraphics{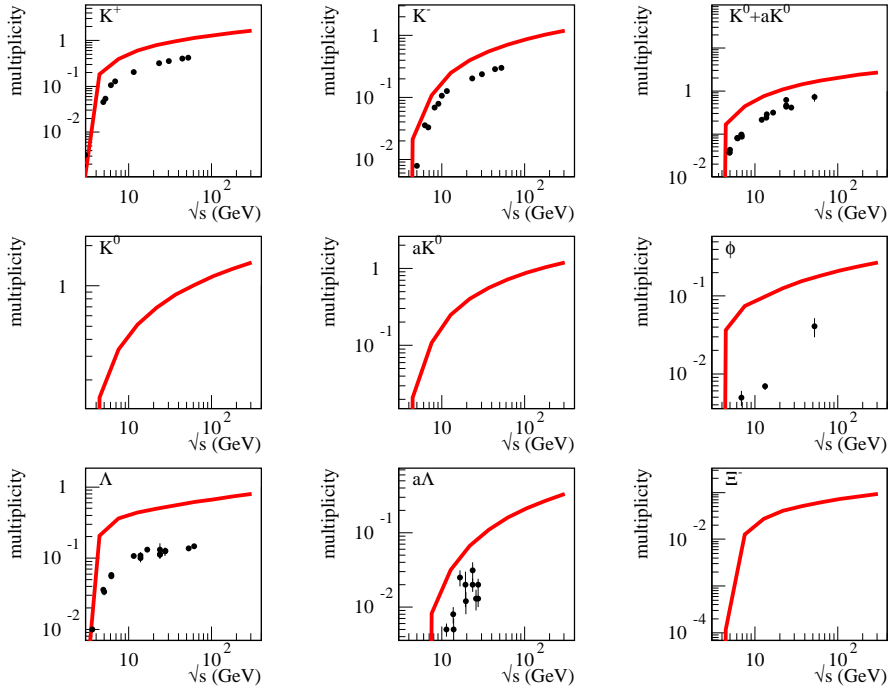}} \par}

\caption{Prediction of the multiplicity of strange hadrons in pp collisions
as a function of \protect\( \sqrt{s}\protect \). The result of the
calculation is compared to the data \cite{rho}.}
\end{figure}

As mentioned above, in the past there has been proposed to use an
additional parameter, \( \gamma _{s} \), in order to describe the
strangeness suppression. This parameter has been interpreted as a
hint that the volume in which strangeness neutrality has to be guaranteed
is small as compared to the volume of the system. However, a detailed
comparison of the multiplicity of all strange particles with the data
shows \cite{bec} that one additional parameter alone is not sufficient
to describe the measured multiplicities of the different strange particles
in a phase space approach to pp collisions.

\section{Transverse Momenta}

Phase space calculations predict not only particle multiplicities
but also the momenta of the produced particles. The average transverse
momenta of the produced particle give a good check whether the energy
density obtained in the fit can really be interpreted as the energy
density of a hadron gas. Fig. 8 shows the average transverse momenta
in comparison with the experimental data \cite{ros}. We see that
over the whole range of beam energies the average transverse momenta
are in good agreement with the data. This confirms that the partition
of the available energy into energy for particle production and kinetic
energy is correctly reproduced in the phase space calculation. It
is remarkable that the average transverse momenta of strange particles
is correctly predicted.
\begin{figure}
{\centering \resizebox*{0.8\textwidth}{!}{\includegraphics{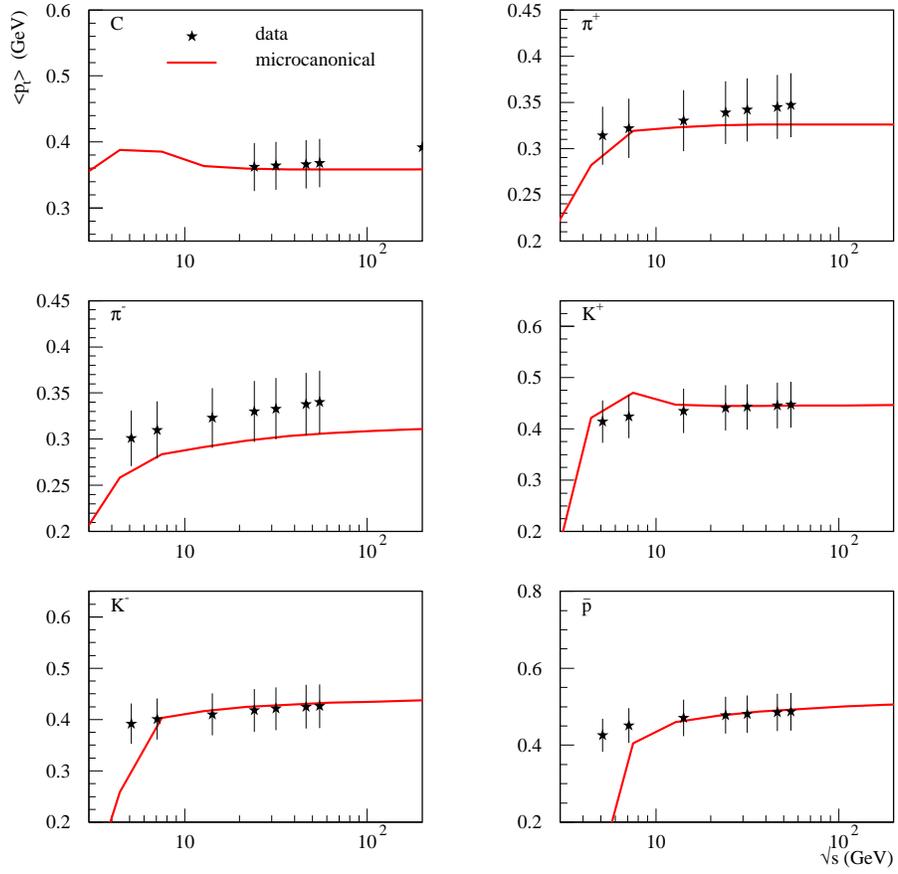}} \par}

\caption{Average transverse momenta as predicted in the phase space calculation
as a function of \protect\( \sqrt{s}\protect \) in comparison with
the experimental values \cite{ros} for different particle species. }
\end{figure}

\section{Energy Fluctuations}

Up to now we have assumed that for a given center of mass energy,
the energy of the droplet \( E \) has an unique value, given in fig.
2. This is of course not a realistic assumption. Most probably the
energy varies from event to event but little is known about the form
of this fluctuation. The only quantity for which data are available
is the multiplicity distribution of charged particles, which has been
the subject of an extensive discussion in the seventies due to the
finding of a scaling law, called KNO scaling. Of course, one can try
now to find an energy distribution which yields the experimental charge
particle distribution but this relation is not unique and therefore
little may be learnt.

It has also been suggested to replace the microcanonical ensemble
calculation, presented here, in favor of a canonical ensemble or an
ensemble where the pressure is the control parameter, however it is
difficult to find a convincing argument. It is the dynamics of the
reaction which determines which fraction of the energy goes into collective
motion, and which fraction into particle production. This has nothing
to do with a heat bath nor with constant pressure on the droplet.
Consequently, the relation between the energy fluctuation, seen in
a system with a fixed temperature, and the true energy fluctuation
is all but evident.

Therefore, we use another approach to study the influence of energy
fluctuations on the observables. We assume that the volume of the
system remains unchanged in order not to have too many variables and
that the energy fluctuates. For technical reasons, we use discrete
distributions, using \( E_{i}=i\, \Delta E \), with \( \Delta E=1 \)GeV.
For \( \sqrt{s}=200\, \mathrm{GeV} \), we have \( \left\langle E\right\rangle =16.15\, \mathrm{GeV} \)
from the above fitting work, and correspondingly we take \( <i>=16.15 \).
We study three cases:\\
 a) The energy distribution is Poissonian \[
\mathrm{Prob}(i)=\frac{\left\langle i\right\rangle ^{i}\mathrm{exp}(-\left\langle i\right\rangle )}{i!}\]
 and\\
 b) The energy distribution is Gaussian \[
\mathrm{Prob}(i)=\left\{ \begin{array}{c}
\frac{1}{0.63}\frac{1}{\sqrt{2\pi }\sigma }\mathrm{exp}(-\frac{E_{i}-\mu }{\sigma })^{2}\, \, \, \, \mathrm{when}\, E_{i}\in [2.5\, \mathrm{GeV},\infty )\\
0\, \, \, \, \, \, \, \, \, \, \, \, \, \, \, \, \, \, \, \, \mathrm{otherwise}.
\end{array}\right. \]
 where an energy threshold of \( 2.5\, \mathrm{GeV} \) is taken for
the proton-proton system. \( \mu =\sigma =14.01\mathrm{GeV} \) to
obtain \( \left\langle i\right\rangle =16.15 \) and the factor 0.63
is used to normalize the energy distribution, and \\
 c) The energy distribution is a negative binomial distribution \[
P^{\mathrm{NB}}(i;\, n,k)=\frac{k(k+1)\cdots (k+i-1)}{i!}\frac{n^{i}k^{k}}{(n+k)^{n+k}}.\]
 The negative binomial distribution is well normalized, and \( <i>=n \).
So we take \( n=16.15 \). The parameter \( k=3 \) is chosen to get
the best fit to multiplicity distribution data from UA5.

All the three types of energy fluctuations are displayed in fig. 9.
\begin{figure}
{\centering \resizebox*{0.4\textwidth}{!}{\includegraphics{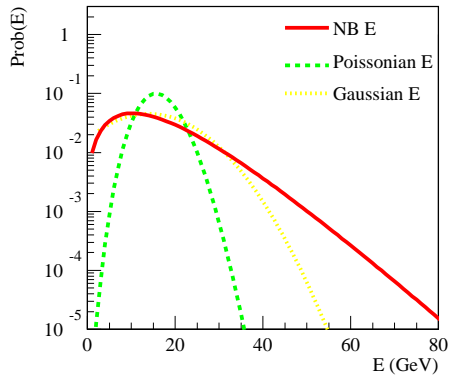}} \par}

\caption{The Gaussian, Poissonian and the NB energy fluctuations. }
\end{figure}
In Fig. 10, we display the influence of these fluctuations on the
charged hadron multiplicity distributions. We compare the results
from a fixed energy of 16.15 GeV, the energy with a fluctuation of
the above-mentioned Poissonian, Gaussian and the NB type. We see that
already for a fixed energy the fluctuation of the charged particle
multiplicity is considerable. Different energy fluctuations give different
multiplicity distributions. The NB energy fluctuation reproduces the
UA5 data for non single diffractive events in antiproton-proton collisions
at \( \sqrt{s}=200\mathrm{GeV} \).
\begin{figure}
{\centering \resizebox*{0.4\textwidth}{!}{\includegraphics{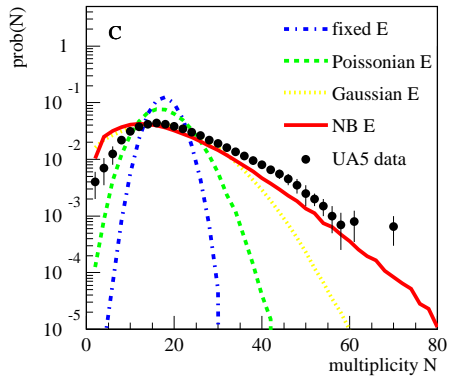}} \par}

\caption{Distribution of the multiplicity of charged hadrons. We display the
results from a fixed energy of 16.15 GeV, the energy with a distribution
of the above-mentioned Poissonian, Gaussian and the 2-NB type. Also
displayed are the UA5 data for non single diffractive events\cite{UA5data}.}
\end{figure}

How does the multiplicity of identified hadrons fluctuate if the droplet
energy fluctuates? This is studied in fig. 11, where we display the
multiplicity distribution of the most copiously produced particles
for fixed energy \( E=16.15\, \, \mathrm{GeV} \) and for a Poissonian,
Gaussian and NB energy distribution. We see here as well that already
for a fixed droplet energy the multiplicity fluctuations are important.
Though different energy fluctuations cause different multiplicity
distribution, the energy fluctuation gives very little effect in the
average multiplicities. So our approach, fixing microcanonical parameters
by fitting the averager multiplicity data, is quite reliable.

There is also a correlation between the pion and kaon multiplicity
in a given event, shown in table.1. The event averaged K/\( \pi  \)
ratio is considerably different from the ratio of the average kaon
and the pion multiplicity. The energy fluctuations change more the
event averaged K/\( \pi  \) ratio than the ratio of the average kaon
and the pion multiplicity.
\begin{figure}
{\centering \resizebox*{0.8\textwidth}{!}{\includegraphics{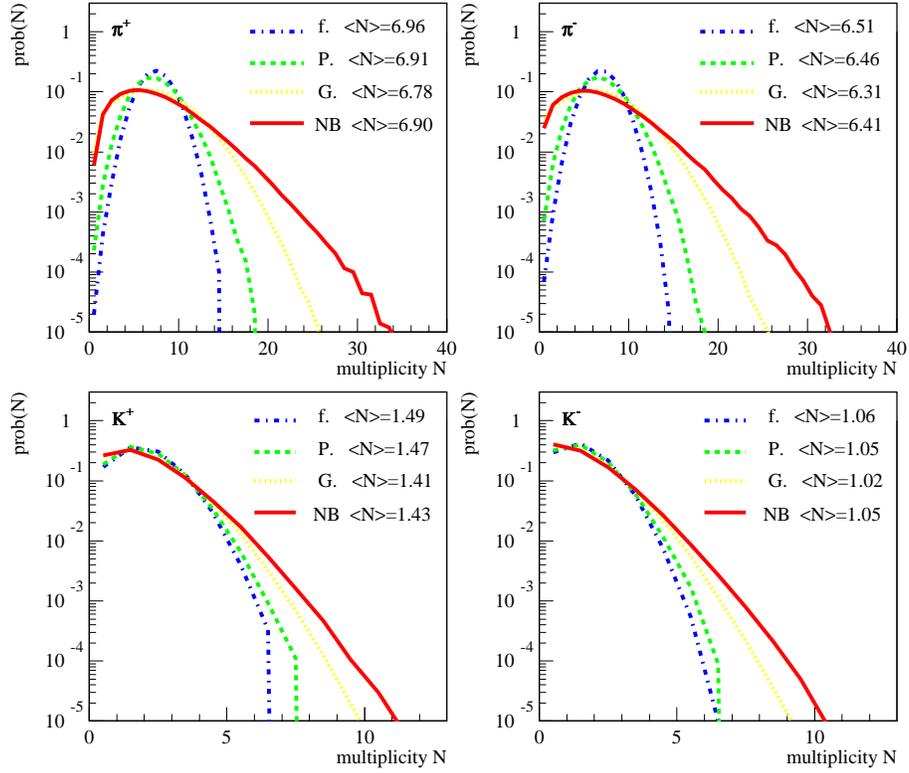}} \par}

\caption{Multiplicity distribution of charged pions and kaons from a fixed
energy of 16.15 GeV, the energy with a distribution of the above-mentioned
Poissonian, Gaussian and the NB. The number refers to the mean multiplicity.}
\end{figure}

\begin{table}
{\centering \begin{tabular}{|c|c|c|c|c|}
\hline 
&
 \( \frac{\left\langle K^{+}\right\rangle }{\left\langle \pi ^{+}\right\rangle } \)&
 \( \frac{\left\langle K^{-}\right\rangle }{\left\langle \pi ^{-}\right\rangle } \)&
 \( \left\langle \frac{K^{+}}{\pi ^{+}}\right\rangle  \)&
 \( \left\langle \frac{K^{-}}{\pi ^{-}}\right\rangle  \)\\
\hline
fixed \emph{E}&
 0.214&
 0.163&
 0.253&
 0.197\\
\hline
Poissonian \emph{E} dis.&
 0.213&
 0.163&
 0.251&
 0.194\\
\hline
Gaussian \emph{E} dis.&
 0.208&
 0.162&
 0.241&
 0.179\\
\hline
NB \emph{E} dis.&
 0.208&
 0.163&
 0.241&
 0.180  \\
\hline
\end{tabular}\par}

\caption{Different K to \protect\( \pi \protect \) ratios where \protect\( \left\langle ...\right\rangle \protect \)
means event averaging.}
\end{table}

\section{Conclusion}

We have presented a micro-canonical phase space calculation to obtain
particle multiplicities and average transverse momenta of particles
produced in pp collisions as a function of \( \sqrt{s} \). Using
the multiplicities of \( p,\, \bar{p},\, \pi ^{+},\, \pi ^{-} \),
we fit the two parameters of the phase space approach, the volume
and the energy.

Using these two parameters, we calculate the multiplicities of all
the other hadrons as well as their average transverse momenta. The
calculated multiplicities agree quite well with experiment as far
as non strange hadrons are concerned.

For the yields of strange hadrons (as well as those with hidden strangeness),
the prediction is off by large factors. In canonical and grand-canonical
approaches, strangeness suppression factor have been used to solve
this problem.

The energy obtained by this fit is much smaller than the energy available
in the center of mass system of the pp reaction, because part of the
energy goes into collective motion in beam direction. Nevertheless,
the average transverse momenta of the produced particles (not only
non-strange but also strange) from this fitting agree quite well with
experiment.

We learn that the volume of the pp collision system increases with
the collision energy. However, it saturates at very high energy(with
Antinnuci's parameterization as input). The maximum value does not
exceed \( 100\, \mathrm{fm}^{3} \). Together with the results from
article \cite{fu}, we conclude that the grand-canonical treatment
cannot describe particle production in pp collisions even at high
energy.

We study the effects from energy fluctuations and find that it is
quite reliable to fix the microcanonical parameters without considering
energy fluctuations.

\noindent \textbf{Acknowledgement}

F.M.L would like to thank H. St\"{o}cker and the theory group in
Frankfurt for the kind hospitality and F. Becattini for many helpful
discussions.


\begin{thebibliography}{10}
\bibitem{Fermi:1950jd}E. Fermi, Prog. Theor. Phys. \textbf{5}, 570 (1950); Phys.Rev. 81
(1951) 683. 
\bibitem{Landau}L. D. Landau, Lzv. Akd. Nauk SSSR 17 (1953) 51; Collected papers of
L. D. Landau, ed. D. Ter Haar, Gordon and Breach, New York, 1965 
\bibitem{hag1}R. Hagedorn, Supplemento al Nuovo Cimento, \textbf{3} (1965) 147.\\
 R. Hagedorn and J. Randt, Supplemento al Nuovo Cimento, \textbf{6}
(1968) 169.\\
 R. Hagedorn, Supplemento al Nuovo Cimento, \textbf{6} (1968) 311. 
\bibitem{Hagedorn}R. Hagedorn, Nucl. Phys. B \textbf{24} (1970) 93. 
\bibitem{Siemens}P. Siemens, J. Kapusta, Phys. Rev. Lett. \textbf{43} (1979) 1486. 
\bibitem{Mekjian}A. Z. Mekjian, Nucl. Phys. A\textbf{384} (1982) 492. 
\bibitem{Csernai}L. Csernai, J. Kapusta Phys. Rep. \textbf{131} (1986) 223. 
\bibitem{Stoecker}H. Stoecker, W. Greiner, Phys. Rep. \textbf{137} (1986) 279. 
\bibitem{Cleymans}J. Cleymanns and H. Satz, Z. Phys. \textbf{C57} (1993) 135; J. Cleymanns,
D. Elliott, H. Satz, and R.L. Thews, CERN-TH-95-298, (1996). 
\bibitem{Rafelski}J. Rafelski and J. Letessier, J. Phys. G: Nucl. Part. Phys. \textbf{28}
(2002) 1819-1832. 
\bibitem{Braun-Munzinger}P.~Braun-Munzinger, K.~Redlich and J.~Stachel, arXiv:nucl-th/0304013. 
\bibitem{bec}F. Becattini and U. Heinz Z.Phys. \textbf{C76} (1997) 269. 
\bibitem{wer}K. Werner and J. Aichelin, Phys. Rev. \textbf{C52} (1995) 1584-1603 
\bibitem{fu}F.M. Liu, K. Werner, J. Aichelin, hep-ph/0304174, Phys. Rev. \textbf{C
68} (2003) 024905. 
\bibitem{gia}M. Antinucci et al., Lett. Nuov. Cim., V6, N4,(1973) 27. \\
 G. Giacomelli, Phys. Rep. \textbf{23C} (1976) 123. 
\bibitem{rho}D. Drijard et al., Z. Phys. \textbf{C9} (1981) 293. 
\bibitem{pp}M. Aiguilar-Benitez et al., Z. Phys. \textbf{C50} (1991) 405. 
\bibitem{ros}R.E. Rossi et al. Nucl. Phys. \textbf{B84} (1975) 269. 
\bibitem{tab}Erhan et al. Phys. Lett. \textbf{85B} (1979) 447. 
\bibitem{UA5data}UA5 Collaboration, R.E. Ansorge \emph{et al.}, Z.Phys.C43:357,1989.\end{thebibliography}
\end{document}